\documentclass[conference]{IEEEtran}

\usepackage{fancyhdr}

\IEEEoverridecommandlockouts
\usepackage{cite}
\usepackage{amsmath,amssymb,amsfonts}
 \DeclareMathOperator{\diag}{diag}

\usepackage{algorithmic}
\usepackage{graphicx}
\usepackage{textcomp}
\usepackage{xcolor}
\def\BibTeX{{\rm B\kern-.05em{\sc i\kern-.025em b}\kern-.08em
    T\kern-.1667em\lower.7ex\hbox{E}\kern-.125emX}}

\usepackage[utf8]{inputenc}
\usepackage[american]{babel}
\usepackage[hidelinks]{hyperref}
\usepackage[capitalize]{cleveref}
\usepackage{mathtools}
\usepackage{booktabs}
\usepackage{fancyhdr}
\usepackage[shortcuts]{extdash}
\usepackage[all]{foreign}
\usepackage{bm}
\usepackage{tikz}
\usetikzlibrary{bayesnet}
\usepackage{siunitx}

\usepackage{caption}
\usepackage{subcaption}
\usepackage{booktabs}
\usepackage{arydshln}
\usepackage{comment}

\usepackage{glossaries}

\newacronym{ap}{AP}{Access Point}
\newacronym{cfr}{CFR}{channel frequency response}
\newacronym{cnn}{CNN}{convolutional neural network}
\newacronym{csi}{CSI}{channel state information}
\newacronym{cv}{CV}{computer vision}
\newacronym{dl}{DL}{deep learning}
\newacronym{edl}{EDL}{Evidential Deep Learning}
\newacronym{har}{HAR}{human activity recognition}
\newacronym{kl}{\mbox{KL}}{Kullback–Leibler}
\newacronym{lan}{LAN}{local-area network}
\newacronym{lstm}{LSTM}{long short-term memory}
\newacronym{mimo}{MIMO}{multiple-input multiple-output}
\newacronym{mlp}{MLP}{multi-layer perceptron}
\newacronym{nic}{NIC}{network interface card}
\newacronym{ood}{OoD}{out-of-distribution}
\newacronym{ofdm}{OFDM}{orthogonal frequency-division multiplexing}
\newacronym{ofdma}{OFDMA}{orthogonal frequency-division multiple access}
\newacronym{phy}{PHY}{Physical Layer}
\newacronym{sdr}{SDR}{software-defined radio}
\newacronym{siso}{SISO}{single-input single-output}
\newacronym{sta}{STA}{station}
\newacronym{vae}{VAE}{Variational Auto-Encoder}
\newacronym{wlan}{WLAN}{wireless local-area network}
\usepackage{amsthm}
\usepackage{amsmath}
\usepackage{amsfonts}
\usepackage{amssymb}

\newcommand{\bb}[1]{\mathbf{#1}}

\newcommand{\bx}{\bb{x}}
\newcommand{\bxi}{\bx^{(i)}}

\newcommand{\bz}{\bb{z}}
\newcommand{\bzi}{\bz^{(i)}}

\newcommand{\bzil}{\bz^{(i,l)}}

\newcommand{\bT}{\boldsymbol{\theta}}

\newcommand{\bphi}{\boldsymbol{\phi}}
\newcommand{\beps}{\boldsymbol{\epsilon}}
\newcommand{\bepsl}{\beps^{(l)}}

\newcommand{\bsigma}{\boldsymbol{\sigma}}
\newcommand{\bmu}{\boldsymbol{\mu}}
\newcommand{\bzero}{\bb{0}}

\newcommand{\bI}{\bb{I}}

\newcommand{\bX}{\bb{X}}

\newcommand{\pT}{p_{\bT}}

\newcommand{\Exp}[2]{\mathbb{E}_{#1}\left[#2\right]}

\newcommand{\eqnr}{\addtocounter{equation}{1}\tag{\theequation}}

\theoremstyle{definition}

\newcommand{\LB}[2]{\mathcal{L}^{#1}(\bT,\bphi; #2)}
\newcommand{\LBT}[2]{\widetilde{\mathcal{L}}^{#1}(\bT,\bphi; #2)}

\usepackage{xspace}

\usepackage[helvratio=0.92]{newtxtext}

\newcommand*\wifi{\mbox{Wi-Fi}\xspace}

\newcommand{\vaef}{\mbox{\sf VAE-F}\xspace}
\newcommand{\vaefIII}{\mbox{\sf VAE-F-3D}\xspace}
\newcommand{\vaeax}{\mbox{\sf VAE-A$x$}\xspace}
\newcommand{\vaeaI}{\mbox{\sf VAE-A1}\xspace}
\newcommand{\vaeaII}{\mbox{\sf VAE-A2}\xspace}
\newcommand{\vaeaIII}{\mbox{\sf VAE-A3}\xspace}
\newcommand{\vaeaIV}{\mbox{\sf VAE-A4}\xspace}

\newcommand{\boldnofusing}[1]{\mbox{\sf\bf No-Fused-{#1}}\xspace}
\newcommand{\boldearlyfusing}{\mbox{\sf\bf Early-Fusing}\xspace}
\newcommand{\boldearlyfusingIII}{\mbox{\sf\bf Early-Fusing-3D}\xspace}
\newcommand{\bolddelayedfusing}{\mbox{\sf\bf Delayed-Fusing}\xspace}

\newcommand{\nofusing}[1]{\mbox{\sf No-Fused-{#1}}\xspace}
\newcommand{\earlyfusing}{\mbox{\sf Early-Fusing}\xspace}
\newcommand{\earlyfusingIII}{\mbox{\sf Early-Fusing-3D}\xspace}
\newcommand{\delayedfusing}{\mbox{\sf Delayed-Fusing}\xspace}

\begin{document}

\title{Accurate Passive Radar via an Uncertainty-Aware Fusion of Wi-Fi Sensing Data
\thanks{
Thanks to Erik Blasch for participating in the project. The work is partially supported by the European Office of Aerospace Research \& Development under award number FA8655-22-1-7017, and by the Air Force Office of Scientific Research under award number FA9550-22-1-0193. Any opinions, findings, and conclusions or recommendations expressed in this material are those of the author(s) and do not necessarily reflect the views of the United States government.}
}

\author{
    \IEEEauthorblockN{
        Marco Cominelli\IEEEauthorrefmark{1},
        Francesco Gringoli\IEEEauthorrefmark{1},
        Lance M. Kaplan\IEEEauthorrefmark{3}, 
        Mani B. Srivastava\IEEEauthorrefmark{4},
        and Federico Cerutti\IEEEauthorrefmark{1}
    }\\
    \IEEEauthorblockA{
        \IEEEauthorrefmark{1} 
            Department of Information Engineering, 
            University of Brescia, Italy. 
            \{name.surname\}@unibs.it
    }
    \IEEEauthorblockA{
        \IEEEauthorrefmark{3}
            DEVCOM Army Research Lab,
            USA.
            lance.m.kaplan.civ@army.mil
    }
    \IEEEauthorblockA{
        \IEEEauthorrefmark{4}
            ECE Department,
            University of California, Los Angeles, USA.
            mbs@ucla.edu
    }
}

\maketitle
\thispagestyle{fancy}

\begin{abstract}
\wifi{} devices can effectively be used as passive radar systems that sense what happens in the surroundings and can even discern human activity.
We propose, for the first time, a principled architecture which employs Variational Auto-Encoders for estimating a latent distribution responsible for generating the data, and Evidential Deep Learning for its ability to sense out-of-distribution activities.
We verify that the fused data processed by different antennas of the same Wi-Fi receiver results in increased accuracy of human activity recognition compared with the most recent benchmarks, while still being informative when facing out-of-distribution samples and enabling semantic interpretation of latent variables in terms of physical phenomena.
The results of this paper are a first contribution toward the ultimate goal of providing a flexible, semantic characterisation of black-swan events, \ie events for which we have limited to no training data.
\end{abstract}

\begin{IEEEkeywords}
 Wireless communications, Machine learning, Sensor fusion
\end{IEEEkeywords}

\section{Introduction}
\label{sec:introduction}
\wifi{} devices can effectively be used as passive radar systems that \emph{sense} what happens in the surroundings and can even discern human activity~\cite{li2020passive}.

In this paper, we propose \--- for the first time \--- a principled analysis of a recently captured dataset of human activities sensed by commercial \wifi{} devices, using \glspl{vae} \cite{kingma_AutoEncodingVariationalBayes_14} for identifying generative relationships with a latent distribution which we use as a compressed view of the original signal.
Specifically, we focus (\cref{sec:motivation}) on the scenario illustrated in \cite{exposingthecsi2023}, where a commercial \wifi{} device performs \gls{har} through the analysis of the \gls{csi} \--- a measurement of the wireless channel's properties \--- detecting the specific activities performed by a target person inside a room.
We evaluate different methods to fuse the data collected simultaneously by different antennas of the same \wifi{} receiver to improve the overall \gls{har} performance.

\glspl{vae} (\cref{sec:generativemodels}) are generative models which can be trained to learn causal relationships (\cf \cref{fig:platevae}) between a latent distribution \--- a multi-variate normal distribution in our case \--- and the training dataset. Differently from standard auto-encoder architectures already employed in analogous \gls{har} tasks \cite{caeradar18}, where the latent space does not have a specific semantic meaning, the \glspl{vae}' assumption that data observations are caused by a latent distribution provides us with a principled method for identifying reasons for the perturbation of the perceived signal.

An uncertainty-aware classifier, notably using the \gls{edl} loss function \cite{Sensoy2018} (\cref{sec:edl}), then utilises the latent distribution of a data sample to identify the actual activity and estimate the \emph{aleatory} (or \emph{aleatoric}) and \emph{epistemic}  uncertainties \cite{cerutti_EvidentialReasoningLearning_22}, hence giving our system the ability to identify \gls{ood} samples (\ie previously unseen activities).

Our experimental results (\cref{sec:results}) \--- which build upon the \delayedfusing \gls{vae}-based architecture we describe in \cref{sec:architecture} \--- support our main hypotheses (\cref{sec:hypothesis}), \ie that \glspl{vae} provide a concise yet informative characterisation of the activities perceived in the \wifi{} signal. In particular, results outperform existing state-of-the-art benchmarks on the same dataset by fusing the sensing of multiple antennas.
Not only does our \delayedfusing  architecture outperforms existing state-of-the-art benchmarks, but it also appears to be informative when facing \gls{ood} samples (\cref{sec:ood}) and provides a semantic interpretation of the \gls{vae} latent distributions through an inherently interpretable model (\cref{sec:dt}).

As we comment in the conclusions (\cref{sec:conclusions}), these results are a first contribution toward the goal of providing a flexible, semantic characterisation of \textit{black-swan} events, \ie events for which there is limited to no training data.

\section{Background}

\subsection{Motivating Scenario}
\label{sec:motivation}
In this paper, we consider an application for \gls{har} in indoor environments using \wifi{} sensing techniques.
Specifically, we focus on the scenario illustrated in \cite{exposingthecsi2023}, where a commercial \wifi{} receiver is used to \textit{sense} the environment through the analysis of the \gls{csi} and detect specific activities performed by an unspecified person inside a room.

\begin{figure}
    \centering
    \includegraphics[width=\columnwidth]{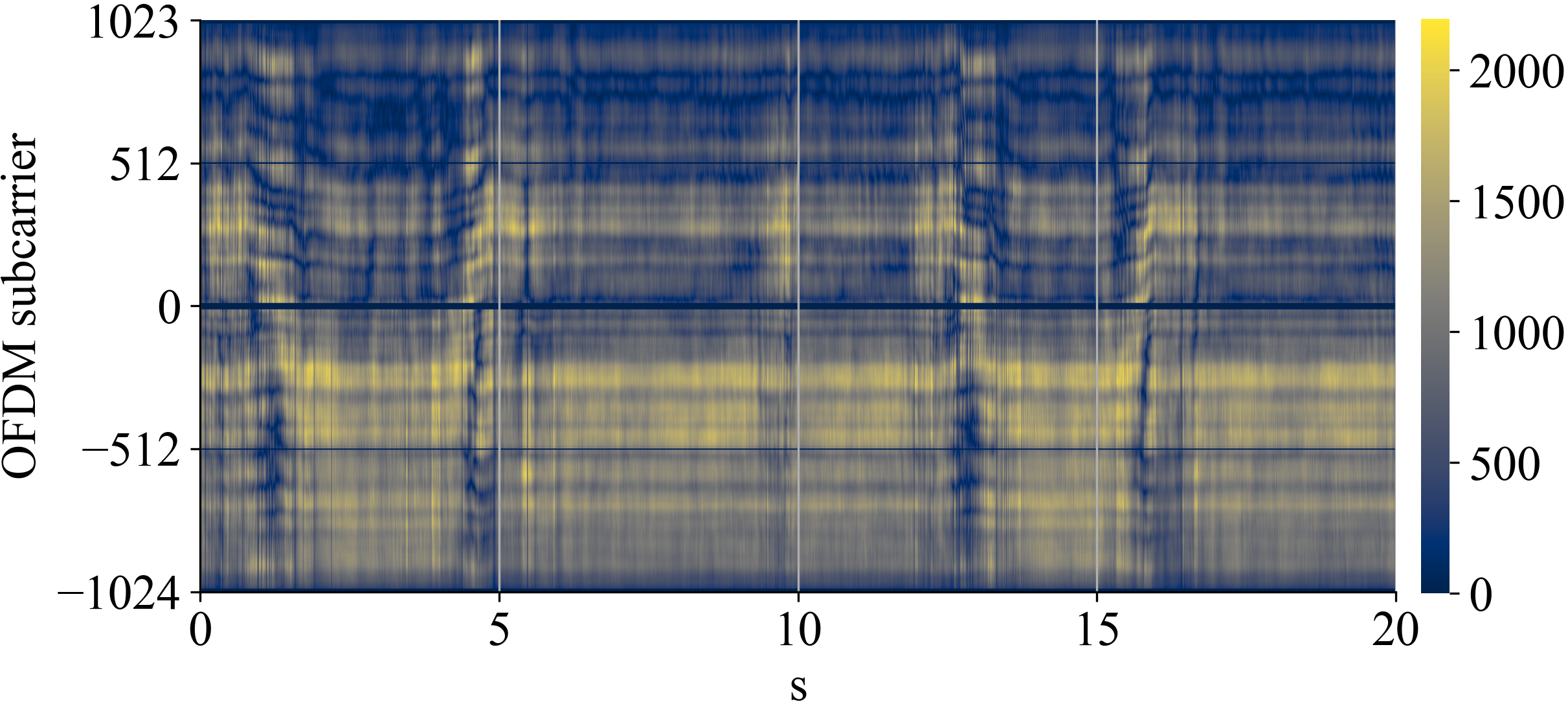}
    \caption{Magnitude of the \acrshort{csi} collected by one antenna while a person is walking. Magnitude values are reported in arbitrary units, as measured inside the \wifi{} chipset.}
    \label{fig:example}
\end{figure}

In wireless communications, the \gls{csi} is an estimation of the wireless channel's properties (see \cref{fig:example}), computed by the receiver for every incoming \wifi{} frame.
The \gls{csi} is a critical element in \gls{ofdm} communication systems (including \wifi{}) because it allows equalising frequency-selective distortions on wide-band communication channels by simply comparing symbols received in the frames' preamble against a known reference signal~\cite{khorov2018tutorial}.
Since such distortions are directly related to the \textit{multipath} effect caused by the physical environment, \gls{csi} analysis is essential for many \wifi{} sensing applications~\cite{ma2019wifi}.
Indeed, the electromagnetic interaction of the wireless signals with the surrounding environment is captured by the \gls{csi} in the form of idiosyncratic interference patterns that depend on the room geometry, the furniture layout, and even the presence and the movements of human bodies.
In other words, the \gls{csi} can be interpreted as an electromagnetic fingerprint of the environment, and \wifi{} receivers can be considered under appropriate assumptions as passive radars.

In this work, we rely on a \gls{csi} dataset recently published for \gls{har}~\cite{exposingthecsi2023}.
The testbed consists of two Asus RT-AX86U routers, each of which has four antennas and has \gls{mimo} capabilities.
During the experiments, one router generates dummy \wifi{} traffic at a constant rate of 150 frames per second using the injection feature of AX-CSI~\cite{axcsi2021}, while the other router (also called \textit{monitor}) uses the same software tool to do the \textit{sensing}, \ie to collect the \gls{csi} from the received test \wifi{} frames.
We consider one single scenario from the dataset in which a person performs different activities in an indoor space of approximately \qty{45}{m^2}.
For every activity, the monitor collects \qty{80}{s} of \gls{csi} data.

\Cref{fig:example} illustrates a snippet of the data captured by one single antenna, \viz the magnitude of the \gls{csi} while a person is walking.
As the person walks around the room, the effect of the environment on the signal changes due to the varying scattering on the human body.
The result is captured in a sort of \emph{spectrogram} that highlights how the relative intensity of the signal changes over time and frequency.
The fundamental assumption of \gls{csi}-based \gls{har} is that it is possible to trace these variations back to the human activity that caused them.
Currently, state-of-the-art \gls{har} systems work by deriving some physically-related quantity from the \gls{csi} that is then used to train a deep learning classification system, like in \cite{meneghello2022,bahadori2022rewis,liu2020}.

In this paper, on the contrary, we evaluate the possibility of using a principled data-driven approach to perform \gls{har}.
The idea of using data-driven approaches on radio data is not new \textit{per se}~\cite{yuan2022}, and an autoencoder-based \gls{har} system has already been proposed in \cite{caeradar18}.
However, we find at least three points that separate our work from the current state of the art.
First, our work builds upon \gls{csi} measurements made with commercial \wifi{} systems, while \cite{caeradar18} employs custom hardware integrated with a \gls{sdr} platform.
Second, by using an \gls{edl} loss function our approach can handle \gls{ood} samples.
Finally, the causal assumptions behind the \gls{vae}-based architecture allow us to (post-hoc) explain the variables in the latent space in terms of physical quantities.

\subsection{Generative Models}
\label{sec:generativemodels}
In this research, we analyse the fundamental dependencies within the \gls{csi} dataset in \cite{exposingthecsi2023} using generative models.
Two prominent families of models can map a datapoint $\bm{x}$ into a class $\mathcal{C}_k$ \cite{Bishop2006}: the \textit{discriminative}, and the \textit{generative}.
Discriminative models estimate the posterior class probabilities $p(\mathcal{C}_k \mid \bm{x})$ and then identify the class with the largest posterior probability.
Generative models, instead, estimate either the class-conditional densities $p(\bm{x} \mid \mathcal{C}_k)$ or directly the joint distribution $p(\bm{x}, \mathcal{C}_k)$.
Generative models can compute the posterior class probabilities and then operate as discriminative models, but they can also generalise to infrequent data points.

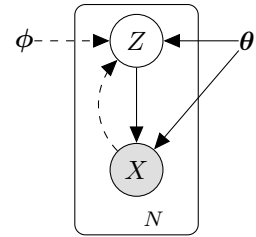
\begin{figure}[t]
\begin{center}
\begin{tikzpicture}
\node[obs] (x1) {$X$};
\node[latent, above=of x1] (z1) {$Z$};
\node[const, left=of z1] (phi1) {$\bm{\phi}$};
\node[const, right=of z1] (theta1) {$\bm{\theta}$};
\edge [dashed] {phi1} {z1};
\edge {theta1} {z1};
\edge {theta1} {x1};
\draw (x1) edge[out=135,in=225,->,dashed] (z1);
\edge {z1} {x1};
\plate [xscale=1.7] {} {(x1)(z1)} {$N$} ;
\end{tikzpicture}
\end{center}
\caption{Plate notation of a \gls{vae} \cite{kingma_AutoEncodingVariationalBayes_14}, with dashed lines denoting the variational approximation. $\bm{\theta}$ are the true yet unknown parameters of the $Z$ distribution which generated the data $X$, while $\bm{\phi}$ are the learnt parameters.}
\label{fig:platevae}
\end{figure}

One of the most popular generative models is the \gls{vae} \cite{kingma_AutoEncodingVariationalBayes_14}, which has become popular for its relatively weak assumptions and fast training via backpropagation.
Let us consider a dataset $\bX = \{\bx^{(i)}\}_{i=1}^N$ consisting of $N$ identical and independently distributed samples of some variable $X$. \Glspl{vae} assume that the data are generated by a random process involving an unobserved continuous random variable $Z$, \cf \cref{fig:platevae}.

As the true parameters $\bT^*$ as well as the values of the latent variables $\bzi$ are unknown,
\gls{vae}s introduce a recognition model whose parameters  $\bm{\phi}$ can be learnt,  $P(Z \mid X, \bm{\phi})$ \--- which is a probabilistic \emph{encoder} that produces a distribution over the possible values of the latent representation $Z$ from which a given datapoint $\bx$ could have been generated \--- which is an approximation to the intractable true posterior $P(Z \mid X)$.
$P(X \mid Z)$ can thus be seen as a probabilistic \emph{decoder} which produces a distribution over the possible values of $\bx$ corresponding to a value of the latent representation $Z$. 

The marginal likelihood is composed of a sum over the marginal likelihoods of individual datapoints $\log p(\bx^{(1)}, \cdots, \bx^{(N)} \mid \bm{\theta}) = \sum_{i=1}^N \log p(\bxi \mid \bm{\theta})$, which can each be rewritten as:
\begin{align}
\begin{split}
\log \pT(\bxi) = & ~\mbox{KL}(p(\bz \mid \bxi, \bm{\phi}) ~||~ p(\bz \mid \bxi, \bm{\theta}))  \\
                 & \qquad + \LB{}{\bxi}
\end{split}
\label{eq:marginallikelihood}
\end{align}
where
$\mbox{KL}(p(\bz \mid \bxi, \bm{\phi}) ~||~ p(\bz \mid \bxi, \bm{\theta}))$
is the (non-negative) \gls{kl} divergence of the approximate from the true posterior
and 
$\LB{}{\bxi}$
is the (variational) \emph{lower bound} on the marginal likelihood of datapoint $i$ \ie:
\begin{align}
\begin{split}
\log \pT(\bxi) & \geq \LB{}{\bxi} = \\
& =  \Exp{p(\bz \mid \bx, \bm{\phi})}{- \log p(\bz \mid \bx, \bm{\phi}) + \log p(\bx,\bz \mid \bm{\theta})}
\end{split}\label{eq:lowerbound}\end{align}
We can thus rewrite \cref{eq:marginallikelihood,eq:lowerbound} as:
\begin{align}
\begin{split}
\LB{}{\bxi} = & - \mbox{KL}(p(\bz \mid \bxi, \bm{\phi}) ~||~ p(\bz \mid \bxi, \bm{\theta})) \\
 & \quad + 
 \Exp{p(\bz \mid \bxi, \bm{\phi})}{\log p(\bxi \mid  \bz, \bm{\theta})}
\end{split}\label{eq:lowerbound2}\end{align}
The chosen approximate posterior $p(\bz \mid \bx, \bm{\phi})$  can be reparametrised (\textit{reparametrisation trick} \cite{kingma_AutoEncodingVariationalBayes_14}) using a differentiable transformation $p(\beps,\bx \mid \bm{\phi})$ with $\beps \sim p(\beps)$.

Often, the \gls{kl} divergence can be integrated analytically, hence only the expected reconstruction error 
$\Exp{p(\bz \mid \bxi, \bm{\phi})}{\log p(\bxi \mid  \bz, \bm{\theta})}$
requires estimation by sampling:
\begin{align}
\begin{split}
\LBT{}{\bxi}
\simeq & - \mbox{KL}(p(\bz \mid \bxi, \bm{\phi}) ~||~ p(\bz \mid \bxi, \bm{\theta})) \\
& \qquad + \frac{1}{L} \sum_{l=1}^L (\log p(\bxi \mid \bzil, \bm{\theta}))
\end{split}\label{eq:estimator2}
\end{align}
where $\bzil = p(\beps,\bx \mid \bm{\phi})$
and
$\bepsl \sim p(\beps)$.
Interpreting \cref{eq:estimator2} as a loss function, we have that the \gls{kl} divergence of the approximate posterior acts as a regulariser, while the second term is an expected negative reconstruction error.

\subsection{Evidential Deep Learning}
\label{sec:edl}
As we will discuss in detail in \cref{sec:methodology}, we aim at investigating the quality of the fusion of different information for the classification of activities from the \gls{csi} dataset. To this end, we wish to rely on a more informative system than a standard discriminative model which returns a categorical distribution over the possible classes. 

\gls{edl} \cite{Sensoy2018} allows us to estimate of the parameters of a Dirichlet distribution $D(\bm{\alpha})$ over the possible $K$ classes. During training, the model pseudo-counts \textit{evidence}, captured by the parameter $\bm{\alpha} \in \mathbb{R}^K$ of the distribution, which is a measure of the amount of support collected from the data in favour of a sample being classified into a particular class.

From this evidence, the belief masses ($b_k$) and uncertainty ($u$) for each class can be calculated as follows. Let $e_{k}\geq0$ be the evidence derived for the $k^{th}$ class:
$b_{k}=\frac{e_{k}}{S}$ and $u=\frac{K}{S}$,
where $K$ is the number of classes and $S=\sum_{i=1}^{K}(e_{i}+1)$, which is the sum of evidence over all classes, is referred to as the Dirichlet strength. We can define the parameters of the output Dirichlet distribution of sample $i$ as $\alpha=f(\mathbf{x}_{i}|\Theta)+1$ where $f(\mathbf{x}_{i}|\Theta)$ represents the evidence vector of sample $i$ given the model parameters. 

During training, the model may discover patterns in the data and generate evidence for specific class labels such that the overall loss is minimized. However, these features may be present in counter-examples, and so reducing evidence may increase the overall loss, despite reducing the loss of these counter-examples. To combat this, a regularisation term is included, which incorporates a \gls{kl} divergence term between a uniform Dirichlet distribution and $\tilde{\alpha}$, where $\tilde{\alpha}$ is the parameters of the output Dirichlet distribution $\alpha$ after removing the non-misleading evidence from $f(\mathbf{x}_{i}|\Theta)$, such that a correctly classified sample with no evidence for other classes will generate $\tilde{\alpha}$ as a uniform Dirichlet distribution.

To learn the parameters $\Theta$ of a neural network, \gls{edl} defines the loss function as
\begin{align}
\begin{split}
    \mathcal{L}(\Theta) = & \sum_{i=1}^{N}\mathcal{L}_{i}(\Theta) \, \\
      & ~ + \lambda_{t}\sum_{i=1}^{N}KL[D(\mathbf{\pi_{i}}|\tilde{\alpha}_{i})\;\|\;D(\mathbf{\pi_{i}}|\langle1,...,1\rangle)]
      \end{split}
\label{equation:loss_kl}
\end{align}
where 
$\lambda_{t}=\min(1.0,t/annealing\_step)\in[0,1]$ is the annealing coefficient, $t$ in the index of the current training epoch, and $annealing\_step$ is the epoch index at which $\lambda_{t}=1$.

Several options for $\mathcal{L}_{i}(\Theta)$ have been considered from \cite{Sensoy2018}, while most of the analysis in the original paper is performed using
\begin{equation*} 
    \mathcal{L}_{i}(\Theta)=\sum_{j=1}^{K}y_{ij}\left(\log(S_{i}) - \log(\alpha_{ij})\right)
\label{equation:edl_log_loss}
\end{equation*}
where $y_{i}$ represents the one-hot vector encoding of the ground-truth label for sample $i$.

\section{Methodology and Hypotheses}
\label{sec:methodology}

\subsection{Architectures}
\label{sec:architecture}
Despite more than a decade of research on \wifi{} sensing, there is still no one-fits-all solution to perform \gls{har} using the \gls{csi}.
In this work, we investigate and compare the performance of several modular architectures that we now present in detail.\footnote{Our dataset and code are available at \url{https://zenodo.org/record/7983057} and at \url{https://github.com/marcocominelli/csi-vae/tree/fusion2023}.}

\begin{figure}
    \centering
    \includegraphics[width=\linewidth]{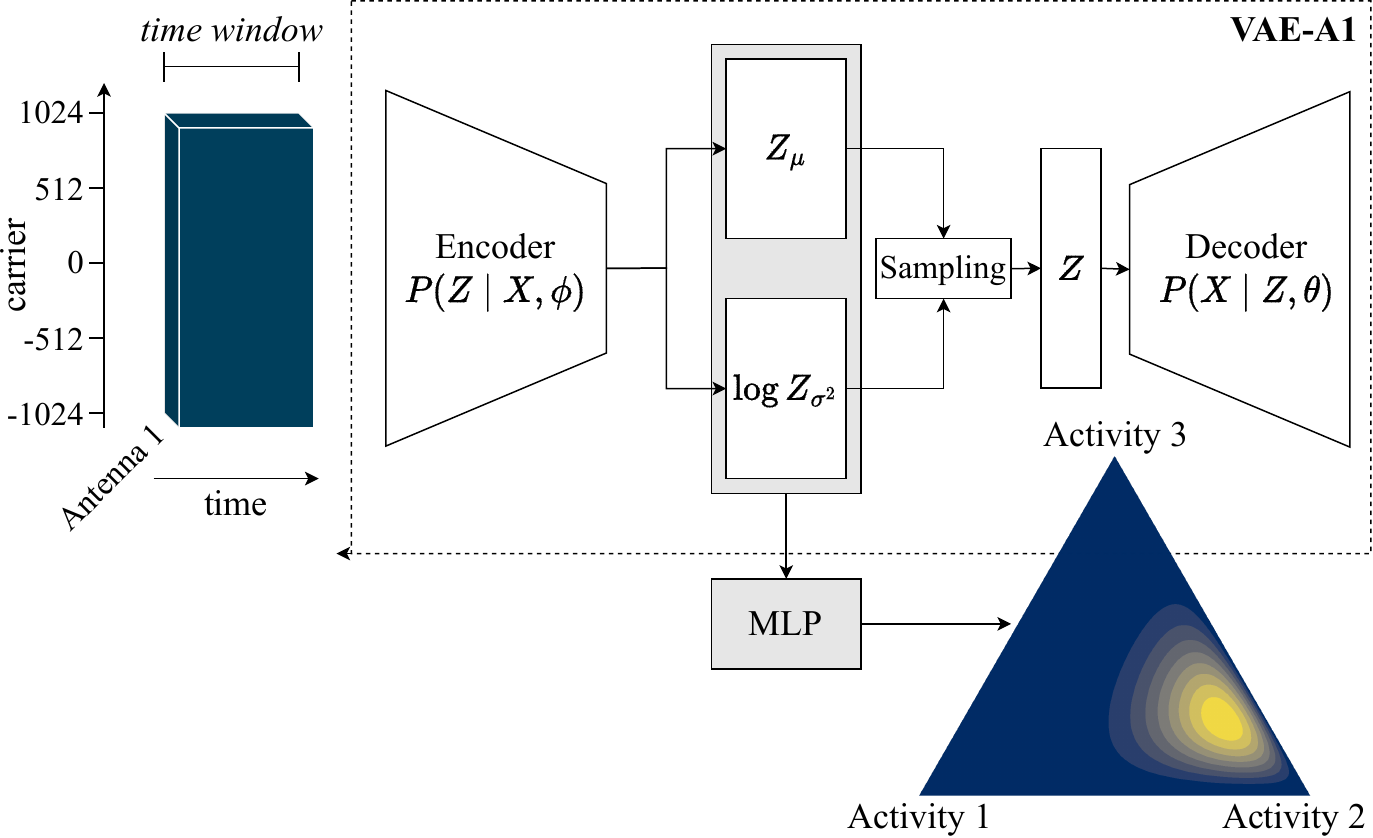}
    \caption{\nofusing{1} architecture. The output of the \acrshort{mlp} is over three activities only for the purpose of showing that it is a Dirichlet distribution. Architectures \nofusing{2}, \nofusing{3}, \nofusing{4} focus on antenna 2, 3, and 4, respectively.}
    \label{fig:nofusingI}
\end{figure}

\begin{figure}
    \centering
    \includegraphics[width=\linewidth]{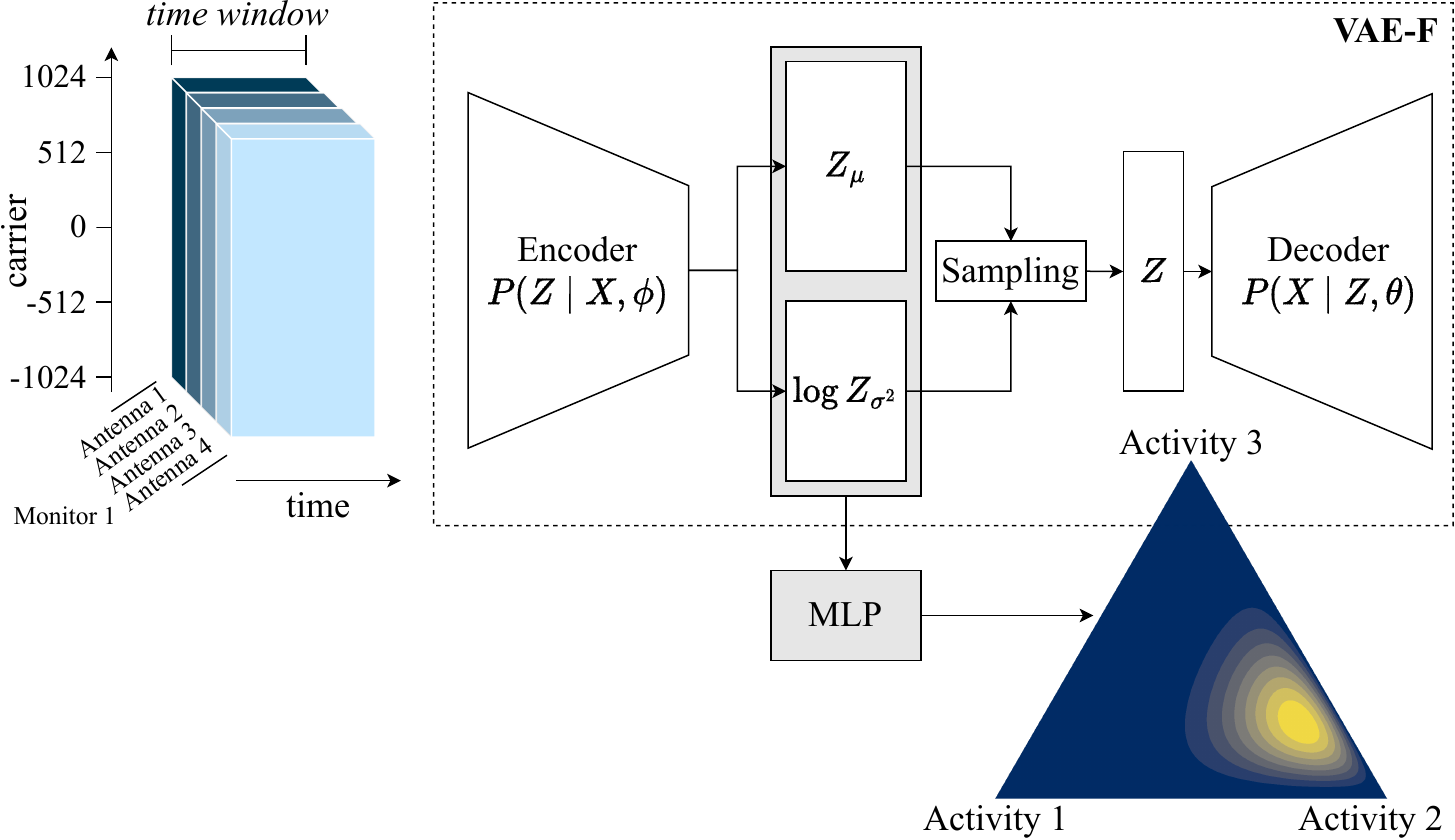}
    \caption{\earlyfusing architecture.  The output of the \acrshort{mlp} is over three activities only for the purpose of showing that it is a Dirichlet distribution.}
    \label{fig:vae-fused}
\end{figure}

\begin{figure}
    \centering
    \includegraphics[width=0.96\linewidth]{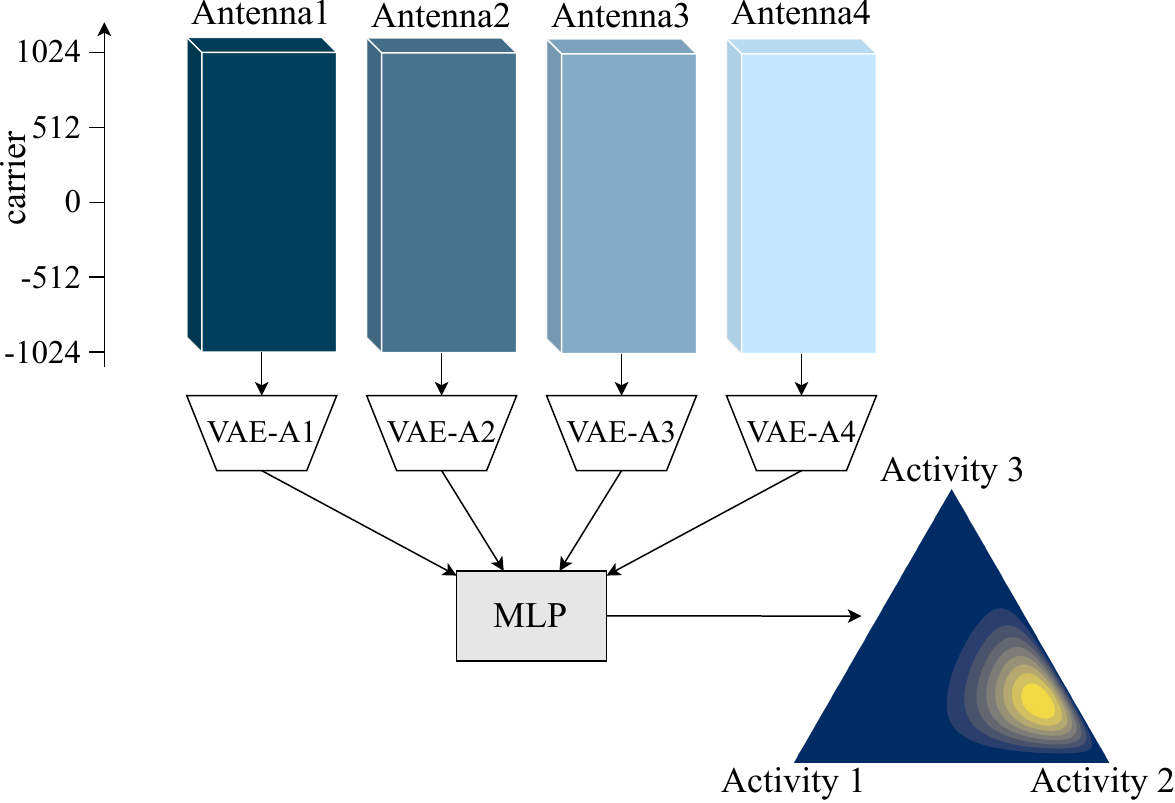}
    \caption{\delayedfusing architecture. The latent space representation of the \acrshort{csi} of every antenna is first extracted from its corresponding \acrshort{vae}, and then fused together at the input of the \acrshort{mlp}. The output of the \acrshort{mlp} is over three activities only for the purpose of showing that it is a Dirichlet distribution.}
    \label{fig:delayedfusing}
\end{figure}

\begin{table}
    \centering
    \caption{Architecture of \vaeax and \vaef. The input (and output) is a tensor of size (450 $\times$ 2048 $\times$ \#antennas). The latent space has size 4 because each normal distribution has 2 parameters. For \vaefIII, the latent space is of size 6.}
    \label{tab:vae_params}
    \begin{tabular}{lccc}
        \toprule
        \textbf{Layer} & \textbf{Kernel size/Nodes} & \textbf{Stride} & \textbf{Activation} \\
        \midrule
            \textit{Encoder} & & & \\
            Conv2D  & $(5,8) \times 32$ & (5,8) & ReLU \\
            Conv2D  & $(5,8) \times 32$ & (5,8) & ReLU \\
            Conv2D  & $(2,4) \times 32$ & (2,4) & ReLU \\
            Flatten & -                 &  -    & -    \\
            Dense   & 16                &  -    & ReLU \\
        \midrule
            Latent space   & dim = 4 & & \\
        \midrule
            \textit{Decoder} & & & \\
            Dense         & 2304              &  -    & ReLU \\
            Reshape       & $(9,8) \times 32$ &  -    & -    \\
            Conv2D$^\top$ & $(2,4) \times 32$ & (2,4) & ReLU \\
            Conv2D$^\top$ & $(5,8) \times 32$ & (5,8) & ReLU \\
            Conv2D$^\top$ & $(5,8) \times 32$ & (5,8) & ReLU \\
        \bottomrule
    \end{tabular}
    \vspace{-0.3cm}
\end{table}

The first set of architectures, called \boldnofusing{$x$} and illustrated in \cref{fig:nofusingI}, includes a \gls{vae} (\vaeax) with a latent bivariate normal distribution trained over the data coming from a single antenna of the \wifi{} monitor \--- \eg the first antenna in \cref{fig:nofusingI} \--- and a \gls{mlp} for classification.
The structure of the \gls{vae} is reported in \cref{tab:vae_params}.
Our monitor contains four antennas, hence we have four architectures trained separately, one for each antenna: \boldnofusing{1}, \boldnofusing{2}, \boldnofusing{3}, and \boldnofusing{4}.

As of today, most of the research on \wifi{} sensing is still based on a single-antenna \gls{csi} extraction system~\cite{csitool2011}.
While many works showed that this kind of data is informative enough for some \gls{har} applications (\eg \cite{guo2019wiar,rfnet2020}), modern \wifi{} systems usually have more than one antenna available.
Therefore, it is worth investigating whether there are any advantages in considering the multiple \gls{csi} data streams to improve \wifi{} sensing performance. 

The second architecture, called \boldearlyfusing and illustrated in \cref{fig:vae-fused}, includes a \gls{vae} (\vaef) with a latent bivariate normal distribution trained over the data coming from the four antennas of the monitor, and a \gls{mlp} for classification.
We also experimented on a variation (\boldearlyfusingIII) of this architecture, where the \gls{vae} considers a latent tri-variate normal distribution (\vaefIII) and then a \gls{mlp} for classification which takes as input the value of the latent space.

The third architecture instead, called \bolddelayedfusing and illustrated in \cref{fig:delayedfusing}, includes the four \glspl{vae} trained on single antennas \--- the same used in \nofusing{1}, \nofusing{2}, \nofusing{3}, \nofusing{4} \--- whose latent spaces are given as input to a single \gls{mlp} for classification.

It is necessary to clarify how we feed the \gls{csi} into the \glspl{vae}.
In general, the \gls{csi} is a complex vector representing the frequency response of the wireless channel (\cf \cref{sec:motivation}).
However, in this work, we discard any information about the phase and we only consider the magnitude of the \gls{csi} for simplicity.
At this point, a sequence of \gls{csi} can be interpreted as a spectrogram, exactly as reported in \cref{fig:example}, which we normalize in amplitude with respect to the highest value measured in the entire dataset.
Instead of feeding the entire spectrogram to the \glspl{vae}, we use a sliding window of \qty{3}{s} to select the input samples for the \gls{vae}.
\vaeaI, \vaeaII, \vaeaIII, and \vaeaIV receive in input the \gls{csi} data from their corresponding antennas.
For \vaef instead, we stack the \gls{csi} of the four antennas along a third dimension, in a way that can be analogously interpreted as four separate channels of a single image.
It is important to notice that this is not the \textit{best} solution for processing the \gls{csi}; on the contrary, we are deliberately removing some of the information.
However, we leave experimentation with different data structures and manipulation techniques as future work.

For each of the \glspl{vae}, we assume that the prior over $Z$ is the centred isotropic multivariate Gaussian $p(\bz) = \mathcal{N}(\bz \mid \bzero, \bI)$, and $P(X 
 \mid Z)$ be a multivariate Gaussian. To approximate the posterior $P(Z \mid X, \bm{\phi})$, let's assume that the true (but intractable) posterior takes on an approximate Gaussian form with approximately diagonal covariance. In this case, we can let the variational approximate posterior be a multivariate Gaussian with a diagonal covariance structure:
\begin{align*}
\log p(\bz 
\mid \bx, \bm{\phi})
&= \log \mathcal{N}(\bz \mid \bmu^{(i)}, \diag(\bsigma^{2 (i)}))
\eqnr\end{align*}
where the mean and standard deviation of the approximate posterior, $\bmu^{(i)}$ and $\bsigma^{(i)}$, are outputs of the encoder.

Let us then sample $L$ values from the posterior $\bzil \sim p(\bz \mid \bxi, \bm{\phi})$ using $\bzil = g(\bxi, \bepsl \mid \bm{\phi}) = \bmu^{(i)} + \bsigma^{(i)} \odot \bepsl$ where $\bepsl \sim \mathcal{N}(\bzero,\bI)$, and $\odot$ signifies an element-wise product (reparameterization trick).

The resulting estimator for this model and datapoint $\bxi$ is thus:
\begin{align}
\begin{split}
\LB{}{\bxi}
\simeq &
\frac{1}{2} \sum_{j=1}^J 1 + \log ((\sigma_j^{(i)})^2) - (\mu_j^{(i)})^2 - (\sigma_j^{(i)})^2  \\
& \quad + \frac{1}{L} \sum_{l=1}^L \log p(\bxi \mid \bzil, \bm{\theta})
\end{split}\label{eq:gaussian_estimator}\end{align}
where 
$\bzil = \bmu^{(i)} + \bsigma^{(i)} \odot \beps^{(l)}$,
$\bepsl \sim \mathcal{N}(0,\bI)$.

Each of the \glspl{mlp} is trained using the EDL loss function discussed in \cref{sec:edl}, hence they output the parameters of a posterior Dirichlet distribution over the possible classes.

\subsection{Experimental Hypotheses}
\label{sec:hypothesis}
We now formulate the hypotheses we will use later while presenting the experimental results (\cf \cref{sec:results}).

We assume the target person can only perform a subset of the possible activities in the dataset in \cite{exposingthecsi2023}.
Specifically, we consider the following five activities: \textit{walk}, \textit{run}, \textit{jump}, \textit{sit}, and \textit{empty room}.
We choose them because i) they are the most basic types of activity conceivable, which have also been considered in related work on \gls{csi}-based \gls{har} (\cf \cite{meneghello2022, bahadori2022rewis}), and ii) we can easily compare the results against the benchmark performance reported in \cite{exposingthecsi2023} over the same five activities.

Our main hypothesis is that the \glspl{vae} in the proposed architectures should provide a concise yet informative characterisation of the activities as perceived by the \wifi{} monitor.
The validity of this hypothesis can be assessed by qualitatively estimating the clustering performance of each \gls{vae}.
We stress the fact that each \gls{vae} operates on the raw data (like the \gls{csi} shown in \cref{fig:example}) without any notion of the semantics around the input data.
When presenting the results in \cref{sec:results}, we will compare the different architectures by measuring the classification accuracy over the \gls{mlp}.

We expect the overall accuracy of the \delayedfusing architecture to be not inferior to the one obtained with the \earlyfusing framework, or with any of the single-antenna architectures \nofusing{1}, \nofusing{2}, \nofusing{3}, and \nofusing{4}.
Since the four antennas of the monitor are spaced by more than half wavelength, the received signals are generally deemed uncorrelated due to the multipath effects.
Hence, there should not be a real advantage in using a single \earlyfusing architecture.
Instead, from an engineering perspective, it is arguably easier to employ four antenna-specific \glspl{vae} rather than trying to extract information from a tensor with four orthogonal components.

\section{Experimental Results}
\label{sec:results}
\subsection{Visualisation of the VAEs space}

\begin{figure*}[t]
    \centering
    \includegraphics[width=\textwidth]{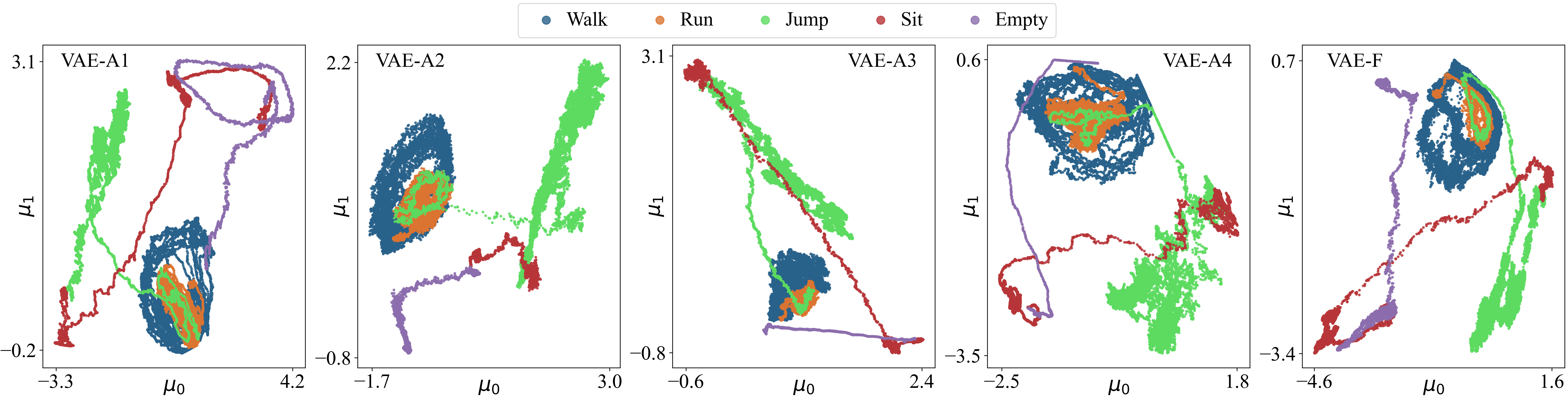}
    \caption{Visualisation of the expected values of the latent space for different \vaeax and \vaef. The encoder maps every \acrshort{csi} onto a point in the $(\mu_0, \mu_1)$ space, coloured as the corresponding activity to help visualise different clusters.}
    \label{fig:vae_latent_space}
    \vspace{-0.3cm}
\end{figure*}

We recall that every architecture introduced in \cref{sec:architecture} is composed of two parts: first, \gls{csi} data are processed by a \gls{vae} that encodes them into a latent representation; then, this representation is used by a \gls{mlp} to detect an activity among a set of target activities.
Therefore, analysing the samples encoded by the \gls{vae} can already give us some insights into the overall performance we can expect from the \gls{har} systems.

In \cref{fig:vae_latent_space}, we represent the expected values of the latent space encoded by different \vaeax (\cf \cref{sec:architecture}).
Each plot shows how different activities are mapped onto the latent space; in particular, every \gls{csi} collected during the corresponding activity is represented by a point in the $(\mu_0, \mu_1)$ space, where $\mu_i$ is the mean of the $i$-th Gaussian variable in the latent space.
We notice that even if the \gls{vae} operates unsupervised, without any notion of the semantics behind the \gls{csi}, it is capable of separating clusters of \gls{csi} data corresponding to different activities.
It is interesting to observe the overlap between the classes \emph{run} and \emph{walk}, with the former that looks like a subset of the latter.
In addition, we observe a partial overlap between classes that share some commonalities, such as \emph{run} and \emph{jump}, or \emph{sit} and \emph{empty room}.

These preliminary results show there are evident differences in the \gls{csi} collected for different activities.
Moreover, at this stage, it looks like there is no clear advantage in fusing the \gls{csi} data together as in the \vaef architecture because the clusters in the latent space look very similar to the ones obtained using just one antenna with \vaeax.

\subsection{Classification Accuracy}
\label{sec:results-accuracy}

The \glspl{mlp} discussed in \cref{sec:architecture} have been implemented as follows with manual parameter tuning:
\begin{itemize}
    \item \nofusing{$x$}, input tensor of dimensions $4 \times 1$, two hidden layers respectively of dimensions $4$ and $8$, both with \textsf{ReLu} as activation function, and an output layer of dimension $5 \times 1$, with activation function \textsf{softplus}, trained over 50 epochs, with batch size $128$, Adam optimiser with learning rate set to $0.01$, and \textit{annealing\_step} $= 22$ (\cf \cref{equation:loss_kl});
    \item \earlyfusing (resp. \earlyfusingIII), input tensor of dimensions $4 \times 1$ (resp. $6 \times 1$), two hidden layers respectively of dimensions $4$ and $8$, both with \textsf{ReLu} as activation function, and an output layer of dimension $5 \times 1$, with activation function \textsf{softplus}, trained over 50 epochs, with batch size $128$, Adam optimiser with learning rate set to $0.001$, and \textit{annealing\_step} $= 22$;
    \item \delayedfusing, input tensor of dimensions $16 \times 1$, two hidden layers respectively of dimensions $16$ and $8$, both with \textsf{ReLu} as activation function, and an output layer of dimension $5 \times 1$, with activation function \textsf{softplus}, trained over 50 epochs, with batch size $128$, Adam optimiser with learning rate set to $0.01$, and \textit{annealing\_step} $= 3$. It is worth noting that choosing this last parameter lower than the one chosen for the other architecture has a detrimental effect on the accuracy, as the penalty induced for misclassified samples in \cref{equation:loss_kl} will be higher. The rationale for this choice will become apparent in \cref{sec:ood}.
\end{itemize}

\begin{figure}
    \centering
    \begin{subfigure}[b]{0.47\linewidth}
         \centering
         \includegraphics[width=\textwidth]{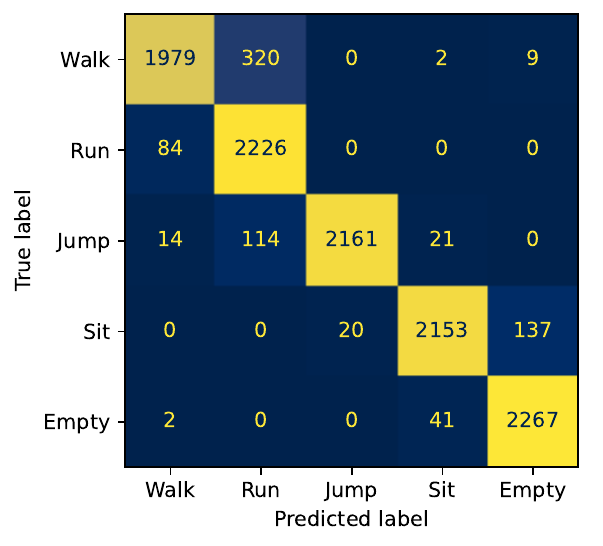}
         \caption{\nofusing{1}}
         \label{fig:cm-no-fused-1}
     \end{subfigure}
    \hfill
    \begin{subfigure}[b]{0.47\linewidth}
         \centering
         \includegraphics[width=\textwidth]{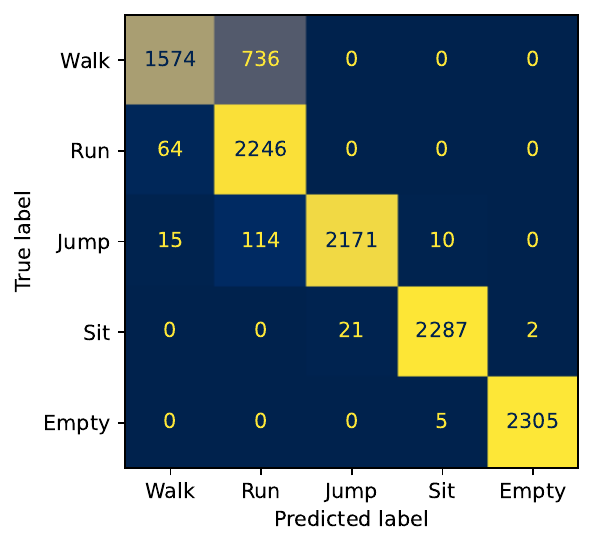}
         \caption{\nofusing{2}}
         \label{fig:cm-no-fused-2}
     \end{subfigure}

    \vspace{0.1cm}
    
    \begin{subfigure}[b]{0.47\linewidth}
         \centering
         \includegraphics[width=\textwidth]{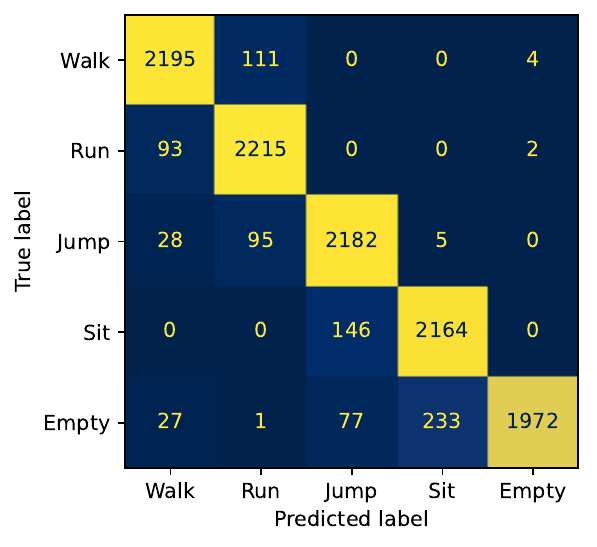}
         \caption{\nofusing{3}}
         \label{fig:cm-no-fused-3}
     \end{subfigure}
    \hfill
    \begin{subfigure}[b]{0.47\linewidth}
         \centering
         \includegraphics[width=\textwidth]{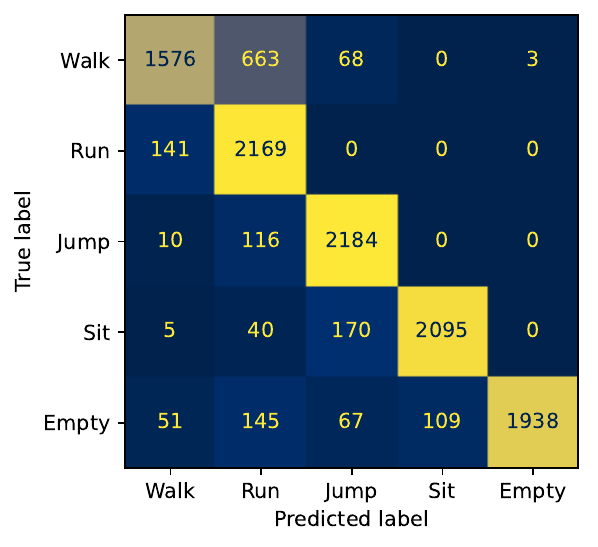}
         \caption{\nofusing{4}}
         \label{fig:cm-no-fused-4}
     \end{subfigure}

    \vspace{0.1cm}
 
    \begin{subfigure}[b]{0.47\linewidth}
         \centering
         \includegraphics[width=\textwidth]{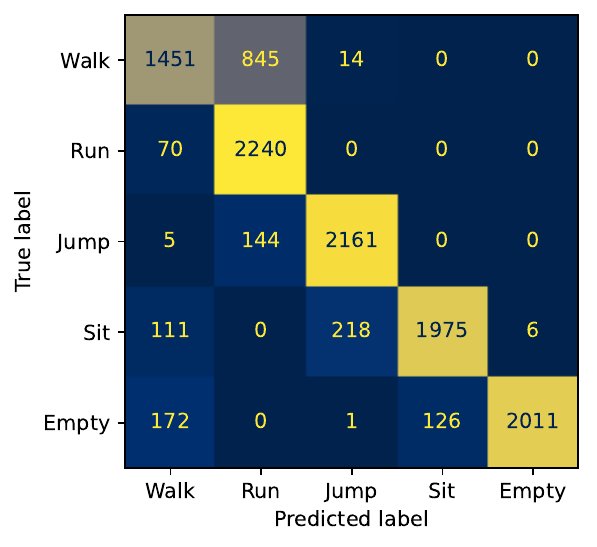}
         \caption{\earlyfusing}
         \label{fig:cm-early-fusing}
     \end{subfigure}
     \hfill
     \begin{subfigure}[b]{0.47\linewidth}
         \centering
         \includegraphics[width=\textwidth]{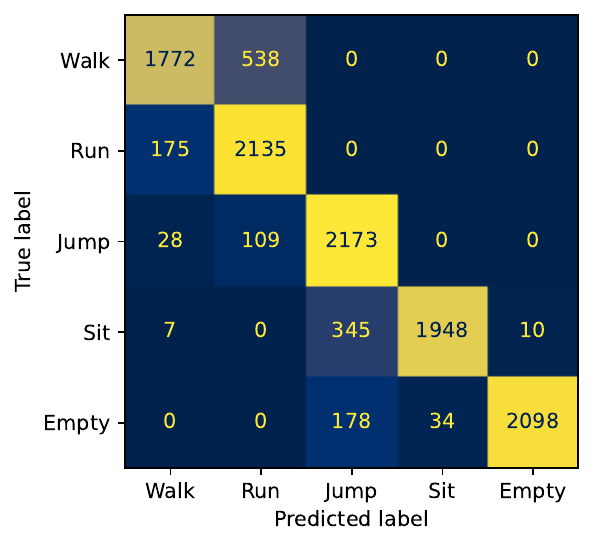}
         \caption{\earlyfusingIII}
         \label{fig:cm-early-fusing3}
     \end{subfigure}
     
    \vspace{0.1cm}
    
    \begin{subfigure}[b]{0.47\linewidth}
         \centering
         \includegraphics[width=\textwidth]{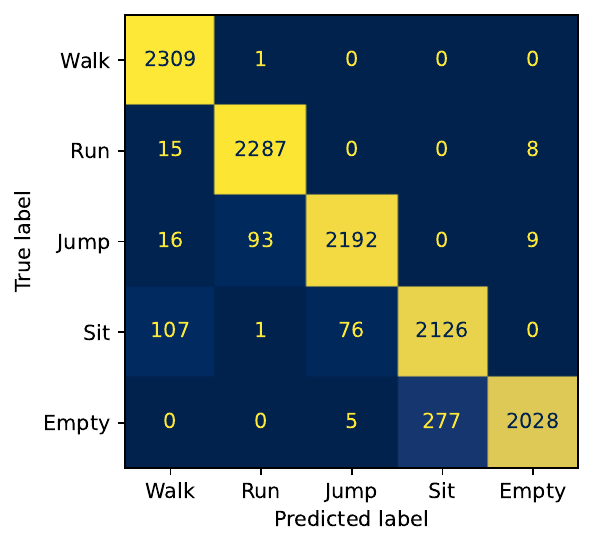}
         \caption{\delayedfusing}
         \label{fig:cm-delayed-fusing}
     \end{subfigure}
    \caption{Confusion matrixes of the tested architectures.} 
    \label{fig:CM-main-results}
\end{figure}

\begin{table}[]
    \centering
    \caption{Performance comparison of the proposed models.}
    \label{tab:main-results}
    {
    \def\arraystretch{1.5}
    \begin{tabular}{l S[table-format=1.2] S[table-format=1.2] S[table-format=1.2] S[table-format=1.2]}
        \toprule
        \textbf{Model} & \textbf{Accuracy} & \textbf{Precision} & \textbf{Recall} & \textbf{F1}\\
        \midrule
            \nofusing{1} & 0.93 & 0.94 & 0.93 & 0.93\\
        \hdashline
            \nofusing{2} & 0.92 & 0.93 & 0.92 & 0.92\\
        \hdashline
            \nofusing{3} & 0.93 & 0.93 & 0.93 & 0.93\\
        \hdashline
            \nofusing{4} & 0.88 & 0.86 & 0.86 & 0.86\\
        \hdashline
            \earlyfusing & 0.85 & 0.87 & 0.85 & 0.85\\
        \hdashline
            \earlyfusingIII & 0.88 & 0.89 & 0.88 & 0.88\\
        \hdashline
            \delayedfusing & \textbf{0.95} & \textbf{0.95} & \textbf{0.95} & \textbf{0.95}\\
        \bottomrule
    \end{tabular}
    }
\end{table}

\Cref{tab:main-results} summarises the achieved results and \cref{fig:CM-main-results} shows the confusion matrixes for each of the architectures considered.

It is germane to see that \textbf{our main experimental hypothesis is confirmed}: fusing data coming from the antennas (\delayedfusing) gives an advantage compared to using single-antenna for activity recognition (\nofusing{$x$}).
Instead, \earlyfusing appears to suffer from a relatively high misclassification rate, especially between the classes \textit{walk} and \textit{run}, a problem that can be partially mitigated by increasing the dimensionality of the latent space (\earlyfusingIII).

These promising results call for an extended investigation of the performance of the proposed architectures.
With respect to the benchmark reported in the original paper using the same dataset~\cite{exposingthecsi2023}, the \delayedfusing architecture shows \textbf{higher accuracy than the baseline} \gls{har} system with four antennas (which stops just below $90\%$).
In this sense, our results indicate that the pre-processing techniques used in state-of-the-art systems may be sub-optimal with respect to activity recognition tasks.
Furthermore, it is interesting to notice that while the benchmark~\cite{exposingthecsi2023} only confuses the classes \textit{walk} and \textit{run}, most of the misclassifications for the \delayedfusing system are between \textit{sit} and \textit{empty} (see \cref{fig:cm-delayed-fusing}).

When dealing with data from a single antenna, the \nofusing{$x$} architectures \textbf{clearly outperform the benchmark} results, achieving an astounding $\mathbf{88\%-93\%}$ (depending on the antenna considered) classification accuracy against the $70\%$ accuracy of the benchmark~\cite{exposingthecsi2023}.
These results suggest that the deterministic pre-processing of \gls{csi} data as proposed in the related work might be sub-optimal or fail to capture useful features of the input data.

While future work could investigate further the early-fusing architectures so as to identify a set of hyperparameters for having higher accuracy, we also point out that \vaef and \vaefIII are architectures substantially more computationally expensive than \vaeax.
Trade-offs will be necessary, but we believe that \delayedfusing is a solution more appealing than \vaef.
It is easier to train, and it has the potential to be deployed in the firmware for every single antenna, thus allowing for greater modularity and reuse.

\section{Post-Hoc Analysis and Discussions}
The \delayedfusing architecture not only outperforms existing state-of-the-art benchmarks (see previous section), but it is informative for \gls{ood} samples, or \textit{black-swan events} (\cref{sec:ood}).
Moreover, we show that we can provide a semantic interpretation of the \gls{vae} latent distributions by using an inherently interpretable model   \cite{burkart_SurveyExplainabilitySupervised_21} (\cref{sec:dt}).

\subsection{Dealing with out-of-distribution samples}
\label{sec:ood}

\gls{edl}'s loss functions (\cref{sec:edl}) can be used to handle epistemic and aleatory uncertainty. In particular, the intuition at the root of \cref{equation:loss_kl} is that ambiguous samples should lead to a uniform distribution, a special case of the Dirichlet distribution with all the parameters equal to 1. 

\begin{figure}
    \centering
    \begin{subfigure}[b]{0.48\linewidth}
         \centering
         \includegraphics[width=\textwidth]{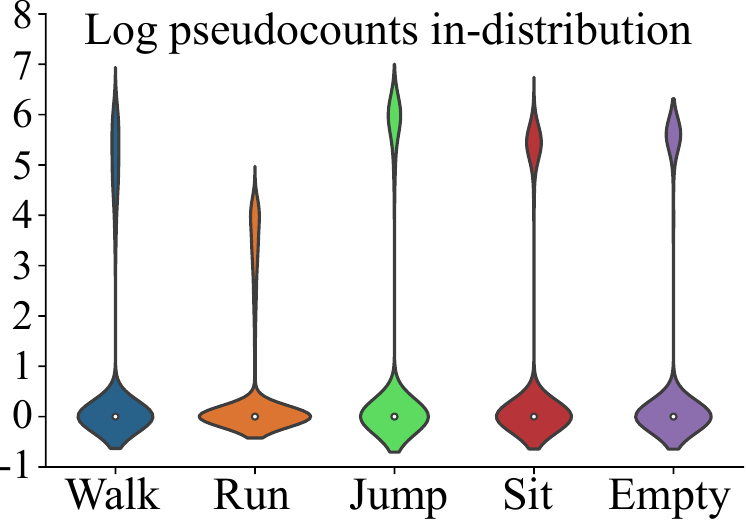}
         \caption{}
         \label{fig:logpseudocountsnosquat}
     \end{subfigure}
    \hfill
    \begin{subfigure}[b]{0.48\linewidth}
         \centering
         \includegraphics[width=\textwidth]{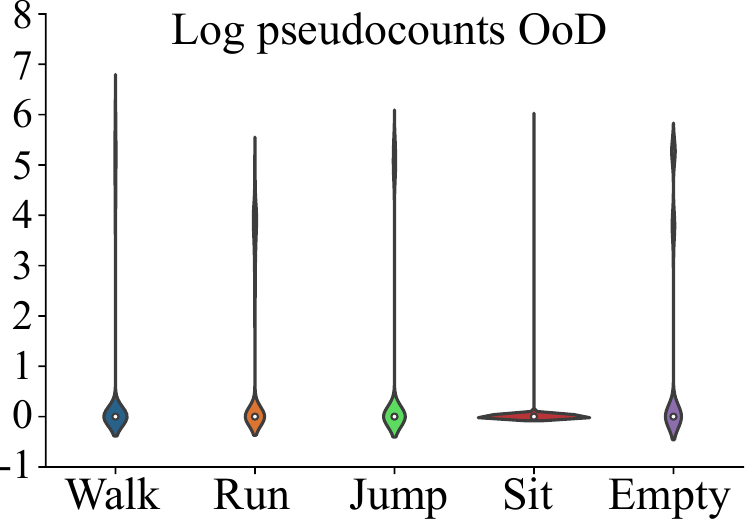}
         \caption{}
         \label{fig:logpseudocountssquat}
     \end{subfigure}
    \caption{Distribution of log-pseudocounts from in-distribution samples (\subref{fig:logpseudocountsnosquat}) and \gls{ood} samples, \ie squatting (\subref{fig:logpseudocountssquat}).}
    \label{fig:ood}
\end{figure}

We simulated this case by considering an activity \--- squatting \--- never used by any model trained so far, whether in training the \glspl{vae} or the \glspl{mlp}: due to space constraints, we consider the \delayedfusing architecture only which has been trained to be more responsive to \gls{ood} (\cf \cref{sec:results-accuracy}) despite the risk of lower accuracy.
\Cref{fig:ood} depicts the distribution of the logarithm of the Dirichlet distribution parameters (pseudo-counts of evidence for each class) outputted by the \gls{mlp} with the \gls{edl} loss function.
When we consider in-distribution samples from the five target activities (\cref{fig:logpseudocountsnosquat}), the resulting distribution has an overall higher Dirichlet strength \--- that is, a higher number of pseudo-counts \--- than when considering \gls{ood} samples (\cref{fig:logpseudocountssquat}).
This means that the proposed architecture effectively handles the uncertainty of the estimation, as we recall that the lower the Dirichlet strength, the closer the output is to the uniform distribution.

In future work, we will expand this preliminary investigation considering more articulated loss functions \cite{cerutti_EvidentialReasoningLearning_22}, including adversarial training with automatically-generated   \cite{sensoy_UncertaintyAwareDeepClassifiers_20}  or explicit \cite{malinin_Predictiveuncertaintyestimation_18} \gls{ood} samples, and more explainable models including probabilistic circuits, recently expanded for dealing with beta distributions \cite{cerutti_HandlingEpistemicAleatory_22} \--- a special case of the Dirichlet distribution. 

\subsection{Physical interpretation of the VAEs' latent space}
\label{sec:dt}

\begin{figure}
    \centering
    \includegraphics[width=\linewidth]{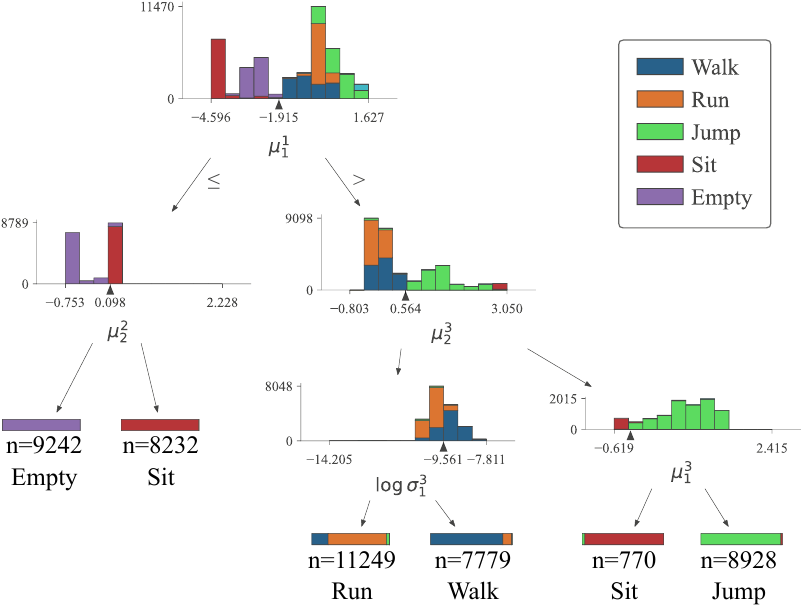}
    \caption{A decision tree with $0.91$ accuracy as an alternative to the \gls{mlp} used in \delayedfusing. $\mu_x^y$ (resp. $\sigma_x^y$) is the expected value (resp. standard deviation) of the $x$-th marginal gaussian of the $Z_y$ latent space.}
    \label{fig:dt}
    \vspace{-0.3cm}
\end{figure}

\begin{figure}
    \centering
    \includegraphics[width=0.48\linewidth]{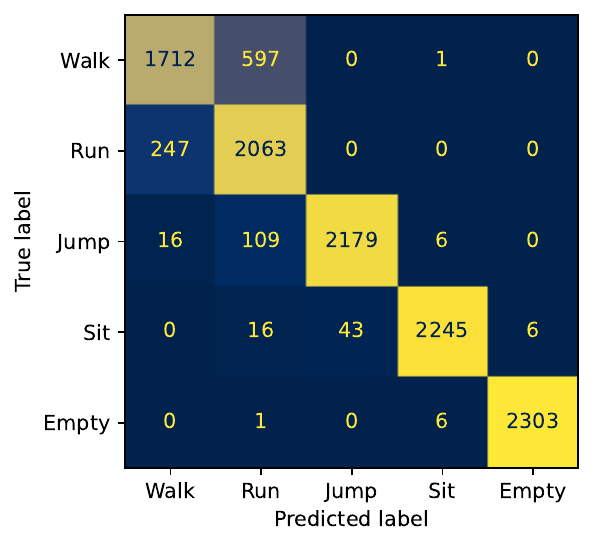}
    \caption{Confusion matrix of the decision tree in \cref{fig:dt}.}
    \label{fig:cmdt}
    \vspace{-0.3cm}
\end{figure}

To provide an interpretation of the \glspl{vae}' latent spaces, let us consider the \delayedfusing architecture where the classification is provided by a decision tree instead of an \gls{mlp} with \gls{edl} loss function. \Cref{fig:dt} illustrates the resulting decision tree, which has an overall test accuracy of $0.91$
despite being specifically designed to have at most three levels; \cref{fig:cmdt} shows the confusion matrix.

By inspecting the resulting decision tree (\cref{fig:dt}), we can build the following working hypotheses concerning the physics captured by the various latent variables. $Z_1$ of the first antenna appears to be particularly sensitive to movements of an individual in the room: the first decision node, indeed, considers the expected value of this variable, and the leaves to the left are all belonging to the classes Empty or Sit. $Z_2$ of the second antenna seems to capture the presence of obstacles to the signal in the environment.
Antenna 3 appears to have embedded the concept of speed, with its ability \--- combining its $Z_1$ and $Z_2$ \--- to distinguish between Run, Walk, and Jump. 

This post-hoc analysis also suggests that not all the human activity considered in this dataset should be treated equally. Walking or running appear to be \textit{atomic} while jumping, with different speeds of movement in the space, could be decomposed into a repeating sequence of \--- perhaps \--- squatting, walking, sitting. To this end, as part of future work, we plan to study \gls{har} combining atomic actions using neuro-symbolic complex event processing systems, \eg \cite{xing_DeepCEPDeepComplex_19, roigvilamala_DeepProbCEPNeurosymbolicApproach_23}.

\section{Conclusion}
\label{sec:conclusions}
We propose, for the first time, a principled architecture which employs \glspl{vae} for identifying causal relationships within a latent distribution and a dataset of human activities measured by passive \wifi radars, and Evidential Deep Learning for its ability to discern out-of-distribution activities.
Not only does our proposed architecture (\cref{sec:architecture}) outperforms existing state-of-the-art benchmarks (\cref{sec:results}), but it can also detect \gls{ood} samples (\cref{sec:ood}).
Moreover, we can provide a semantic interpretation of the \gls{vae} latent distributions by using an interpretable model (\cref{sec:dt}).

The results of this paper are a first contribution toward the goal of a flexible, semantic characterisation of \textit{black-swan events}, \ie events for which we have limited to no training data.
As part of future work, indeed, we aim at employing more sophisticated methodologies for estimating epistemic and aleatory uncertainties, \eg \cite{sensoy_UncertaintyAwareDeepClassifiers_20}, and for reasoning about them, \eg \cite{cerutti_HandlingEpistemicAleatory_22}, building upon also recent advancements in neuro-symbolic learning and reasoning, \eg \cite{xing_DeepCEPDeepComplex_19, roigvilamala_DeepProbCEPNeurosymbolicApproach_23}.
Future challenges include studying the ability of the proposed architecture to generalise to different people performing other activities in various environments~\cite{exposingthecsi2023}, as well as different applications that make use of radar technology, \eg \cite{braca2022}.

\bibliographystyle{IEEEtran} 
\bibliography{csipaper, references, biblio}
\end{document}